\documentclass[12pt]{article}
\usepackage{latexsym,amsfonts,amssymb}
\makeatletter
\@addtoreset{equation}{subsection}
\makeatother

\begin{document}
\begin{center}
{\Large \bf Compensational Gravity\\
Fundamentals and an Application: The Cycling Universe}
\\[1.5cm]
 {\bf Vladimir S.~MASHKEVICH}\footnote {E-mail:
  Vladimir.Mashkevich100@qc.cuny.edu}
\\[1.4cm] {\it Physics Department
 \\ Queens College\\ The City University of New York\\
 65-30 Kissena Boulevard\\ Flushing, New York
 11367-1519} \\[1.4cm] \vskip 1cm

{\large \bf Abstract}
\end{center}
Compensational gravity, which is regarded as a fundamental theory, is an advanced version of semiclassical gravity. It is a construction which extends the Einstein equation. Along with the energy-momentum tensor, the extended Einstein equation includes the compensation tensor, or compenson. The latter compensates for the energy-momentum tensor insufficiency, which consists in the discontinuity in time (due to quantum state reduction) and in space (due to sharp cutoff), as well as in an anomaly (nonrealistic state equation and nonzero divergence). The compenson is a primary object, for which equations are formulated. Specifically, purely dark objects may or may not exist. The dynamics of compensational gravity gives rise naturally to the cosmological constant, or dark energy and to dark matter: The compenson versus particle dark matter. On the basis of the dynamics, a cycling model of the closed universe is constructed.
\newpage
\section*{Introduction}

The status of semiclassical gravity is a subject of speculations [1--7]. It is quite natural to put forward semiclassical gravity---a version of quantum gravity, i,e., a unification of quantum theory and general relativity---as a fundamental theory. But the original formulation of the theory, which may be called plain semiclassical gravity, has proved to be inconsistent.

Plain semiclassical gravity was advanced by M{\o}ller [8] and Rosenfeld [9,10]. It is based on the Einstein equation in which the source of gravitational field is represented by the expectation value of the energy-momentum tensor operator. This approach suffers  at least from the following defects: (i) The expectation value refers to a set of results rather than to a specific event [11]; (ii) Quantum state reduction results in the discontinuity of the expectation value of the energy-momentum tensor and, hence, in the discontinuity of the Einstein tensor [12].

To remove the defect (i), we may interpret the source as an effective, or actual value rather than the expectation, or mean value.

As to the defect (ii), the Einstein equation should be extended by inserting a compensation tensor field, or compenson. This has been done in [13,14] with the tensor of a specific form.

In this paper, a general treatment of the compensation concept is presented. Compensation is ensured both for the discontinuity of the effective value of the energy-momentum tensor in time (due to quantum state reduction) and in space (due to sharp cutoff) and for the anomaly of the effective value (nonrealistic state equation and nonzero divergence). The compenson is a primary object, for which equations are formulated.

Pseudoclassical compensational gravity is a theory with compensation in which the quantum origin of the source of the gravitational field is taken into consideration phenomenologically.

The dynamics of compensational gravity gives rise naturally to the cosmological constant, or dark energy and to dark matter. This results in an opposition: The compenson versus particle dark matter.

Specifically, purely dark objects may or may not exist.

The cycling, or oscillating model of the universe is attractive because it avoids the problem of Genesis [15]. The problem of the cycling universe is the subject of a considerable literature [16,15,17--19].

In this paper, we construct a cycling model of the closed universe based on the dynamics of compensational gravity.

\section{Plain semiclassical gravity and its inconsistency}

\subsection{Plain semiclassical gravity}

Plain semiclassical gravity is based on the following equations:

The Einstein equation
\begin{equation} 
G_{\mu\nu}-\Lambda g_{\mu\nu}=8\pi\varkappa T_{\mu\nu}\,,\quad \mu,\nu=0,1,2,3
\end{equation}
where $G_{\mu\nu}$ is the Einstein tensor, $g_{\mu\nu}$ is metric, $\Lambda$ is the cosmological constant, $\varkappa=t^{2}_{\mathrm{Planck}}$ is the gravitational constant, $c=\hbar=1$,
\begin{equation} 
T_{\mu\nu}=(\Psi,\hat{T}_{\mu\nu}\Psi)
\end{equation}
$\hat{T}_{\mu\nu}$ is the energy-momentum operator, and $\Psi$ is a state vector;

The Schr\"odinger equation
\begin{equation} 
\frac{\mathrm{d}\Psi}{\mathrm{d}t}=-\mathrm{i}\hat{H}\Psi\,,\quad t=x^{0}
\end{equation}

\subsection{The interpretation of the source of the gravitational field}

The source of the gravitational field, i.e., $T_{\mu\nu}$ is interpreted as the expectation, or mean value. But that interpretation refers to a set of results, whereas the left side of the Einstein equation relates to a specific point of spacetime.

\subsection{Discontinuity in time}

The reduction of the state vector $\Psi$ results in a discontinuity of $T_{\mu\nu}$ in time, specifically of the components $T_{0\nu}\,,\;\nu=0,1,2,3$. This entails the discontinuity of the derivatives $g_{ij,0}=\partial g_{ij}/\partial t\,,\;i,j=1,2,3$, in the components $G_{0\nu}\,,\;\nu=0,1,2,3$, which, in turn, gives rise to $\delta$-functions in the components $G_{ij}$ .

\subsection{Discontinuity in space}

A spatial sharp cutoff in $T_{\mu\nu}$ entails a discontinuity of the first derivatives of metric components, which results in the appearance of $\delta$-functions in the components of the Einstein tensor.

\subsection{State equation anomaly}

In the case of a phenomenological treatment of $T_{\mu\nu}$, a solution to the Einstein equation (1.1.1) may imply a nonrealistic state equation.

\subsection{Divergence anomaly}

Generally, it is possible that the divergence of the tensor $T_{\mu\nu}$ (1.1.2) is nonzero,
\begin{equation} 
T^{\mu\nu}{}_{;\nu}\neq 0
\end{equation}
whereas
\begin{equation} 
(G-\Lambda g)^{\mu\nu}{}_{;\nu}\equiv 0
\end{equation}

\section{Compensation: Starting statements}

\subsection{Effective value of source}

To remove the defect relating to the interpretation of the expression (1.1.2) for $T_{\mu\nu}$\,, we change the interpretation: $T_{\mu\nu}$ is treated as the effective, or actual value of the energy-momentum tensor rather than the expectation, or mean value.

\subsection{Compenson}

To cope with the discontinuity and anomaly problems, it is necessary to introduce a compensational tensor field, or compenson, $\Theta_{\mu\nu}$\,, which compensates for the insufficiency of the the energy-momentum tensor, $T_{\mu\nu}$:
\begin{equation} 
T_{\mu\nu}\rightarrow T_{\mathrm{compensated}\,\mu\nu}:=T_{\mu\nu}+\Theta_{\mu\nu}\,,\quad \mu,\nu=0,1,2,3
\end{equation}
where $T_{\mathrm{compensated}\,\mu\nu}$ is the compensated, or extended energy-momentum tensor.

The compenson is regarded as a primary object, so that we have to introduce equations for it.

Gravity with compensation in which the quantum origin of $T_{\mu\nu}$ is treated phenomenologically, i.e., without using (1.1.2), (1.1.3), may be called pseudoclassical compensational gravity.

\subsection{The extended Einstein equation}

Under the change (2.2.1), we obtain the extended Einstein equation
\begin{equation} 
G_{\mu\nu}-\Lambda g_{\mu\nu}=8\pi\varkappa T_{\mathrm{compensated}\,\mu\nu}
\end{equation}

\subsection{The conservation laws}

The Bianchi identity
\begin{equation} 
(G-\Lambda g)^{\nu}_{\mu;\nu}\equiv 0
\end{equation}
and (2.3.1) imply the conservation laws, or the equations of motion [20];
\begin{equation} 
(T_{\mathrm{compensated}}[g_{\mu\nu}])^{\sigma}_{\lambda;\sigma}\equiv 0\quad\mathrm{with\;respect\;to}\;g_{\mu\nu}
\end{equation}
which must hold whether or not the Einstein equation is satisfied [21]; (2.4.2) are not equations for metric.

\section{Spacetime manifold and metric}

\subsection{Direct product manifold}

We assume that the spacetime manifold is the direct product:
\begin{equation} 
M^{4}=T\times S\,,\quad M^{4}\ni p=(t\in T,s\in S)
\end{equation}
where $T$ is a 1-dimensional time manifold and $S$ is a 3-dimensional space manifold. The tangent space is the direct sum:
\begin{equation} 
M^{4}_{p}=T_{t}\oplus S_{s}
\end{equation}

\subsection{1+3 metric}

We set
\begin{equation} 
T_{t}\perp S_{s}
\end{equation}
so that metric is of the 1+3 form
\begin{equation} 
\mathrm{d}s^{2}=g_{00}(\mathrm{d}x^{0})^{2}+g_{ij}\mathrm{d}x^{i}\mathrm{d}x^{j}\,,
\quad x^{0}=t\in T,\;(x^{i})\in S
\end{equation}
This metric relates to both a synchronous reference frame and a static one; in the former case \begin{equation} 
g_{00}=1,\;\;g_{ij}=g_{ij}(t,(x^{l}))
\end{equation}
in the latter
\begin{equation} 
g_{00}=g_{00}(x^{l}),\;\;g_{ij}=g_{ij}(x^{l})
\end{equation}

\section{Compensation for the insufficiency \\of the energy-momentum tensor}

\subsection{Compensation for divergence anomaly}

Equation (2.4.2)
\begin{equation} 
\{(T+\Theta)[g_{\mu\nu}])\}^{\sigma}_{\lambda;\sigma}\equiv 0\quad\mathrm{with\;respect\;to}\;g_{\mu\nu}
\end{equation}
resolves the divergence anomaly problem

\subsection{Compensation for discontinuity in time}

The problem of compensation for discontinuity in time which is due to quantum state reduction relates to equations
\begin{equation} 
(G-\Lambda g)_{\mu}^{0}=8\pi\varkappa(T+\Theta)_{\mu}^{0}, \quad\mu=0,1,2,3
\end{equation}
Thus, we come to the conditions:
\begin{equation} 
(T+\Theta)_{\mu}^{0}, \quad\mu=0,1,2,3,\quad\mathrm{are\; continuous\; under\; reduction}
\end{equation}

Generally, in a region with matter,
\begin{equation} 
\Theta_{\mu}^{\nu}\neq 0
\end{equation}

\subsection{Compensation for discontinuity in space}

Consider a boundary surface $\Sigma$ of a material object. Introduce a coordinate system with the metric of the form [22--24]
\begin{equation} 
\mathrm{d}s^{2}=\mathrm{d}s^{2}_{\Sigma}=g_{\Sigma 44}(\mathrm{d}x_{\Sigma}^{4})^{2}+
g_{\Sigma ab}\mathrm{d}x_{\Sigma}^{a}\mathrm{d}x_{\Sigma}^{b},\quad a,b,=1,2,3
\end{equation}
on both sides of the surface $\Sigma$ defined by the equation
\begin{equation} 
x_{\Sigma}^{4}=\mathrm{const}
\end{equation}
The components $g_{\Sigma ab}$ and all derivatives $g_{\Sigma ab,a'}\,,\;g_{\Sigma ab,a'b'}$ should be continuous on the surface.

The matching conditions on the surface $\Sigma$ result in the equations [22]
\begin{equation} 
[\Theta_{\mathrm{external}\Sigma}]_{4}^{4}=[(T+\Theta)_{\mathrm{internal}\Sigma}]_{4}^{4}
\end{equation}
\begin{equation} 
|g_{\mathrm{external}\Sigma 44}|^{1/2}[\Theta_{\mathrm{external}\Sigma}]_{a}^{4}=|g_{\mathrm{internal}\Sigma 44}|^{1/2}[(T+\Theta)_{\mathrm{internal}\Sigma }]_{a}^{4}\,,\quad a=1,2,3
\end{equation}

Now, we have the relations
\begin{equation} 
A_{\mu}^{\nu}=
\frac{\partial x^{\alpha}_{\Sigma}}{\partial x^{\mu}}\frac{\partial x^{\nu}}{\partial x^{\beta}_{\Sigma}}
A_{\alpha}^{\beta},\quad \alpha,\beta=4,1,2,3,\quad \mu,\nu=0,1,2,3,
\quad\mathrm{on}\;\Sigma,\quad A=T,\Theta
\end{equation}

\subsection{Compensation for state equation anomaly}

The extended Einstein equations are of the form
\begin{equation} 
G_{\mu}^{\nu}-\Lambda g_{\mu}^{\nu}=8\pi\varkappa T_{\mathrm{compensated}}{}_{\mu}^{\nu}
\end{equation}

There can be solutions of these equations with $T_{\mathrm{compensated}}{}_{\mu}^{\nu}$ which do not correspond to any realistic state equation. However, $T_{\mu}^{\nu}$ may correspond to such an equation, and
\begin{equation} 
\Theta_{\mu}^{\nu}=T_{\mathrm{compensated}}{}_{\mu}^{\nu}-T_{\mu}^{\nu}
\end{equation}
ensures the compensation.

\section{Dynamics}

\subsection{Synchronous reference frame.\\ Energy, momentum, and stress compensons}
In dynamics, the reference frame is synchronous, so that metric is
\begin{equation} 
\mathrm{d}s^{2}=\mathrm{d}t^{2}+g_{ij}(t,(x^{l}))\mathrm{d}x^{i}\mathrm{d}x^{j},\quad t=x^{0}\,,\quad i,j=1,2,3
\end{equation}

The Christoffel symbols meet the conditions [25]
\begin{equation} 
\Gamma_{00}^{\mu}=0,\;\;\Gamma^{0}_{0\mu}=0,
\quad \mu=0,1,2,3
\end{equation}
i.e.,
\begin{equation} 
\Gamma^{0}_{00}=\Gamma^{i}_{00}=\Gamma^{0}_{0i}=0,\quad i=1,2,3
\end{equation}

Consider a tensor
\begin{equation} 
A_{\mu}^{\nu}=(A_{0}^{0},A_{i}^{0},A_{i}^{j})
\end{equation}
With respect to $(x^{l})$, $A_{0}^{0}$ is a scalar, $A_{i}^{0}$ is a vector, and $A_{i}^{j}$ is a tensor. The compenson is
\begin{equation} 
\Theta_{\mu}^{\nu}=(\Theta_{0}^{0},\Theta_{i}^{0},\Theta_{i}^{j})
\end{equation}
where $\Theta_{0}^{0}$ is the energy compenson, $\Theta_{i}^{0}$ is the momentum compenson, and $\Theta_{i}^{j}$ is the stress compenson.

In compensational gravity, it is assumed that under quantum state reduction the discontinuity of $T_{\mu}^{0}$ in time is compensated by $\Theta_{\mu}^{0}$, i.e., by the energy and momentum compensons. On the other hand, the construction admits of the time discontinuity of $T_{i}^{j}$ without compensation. The compenson $\Theta_{\mu}^{\nu}$ compensates for the discontinuity of $T_{\mu}^{\nu}$ in space.

Now we have to establish dynamical equations for the energy, momentum, and stress compensons.

\subsection{Dynamical equations for the momentum and energy\\ compensons}

From (2.4.2) follows
\begin{equation} 
\Theta^{\nu}_{\mu;\nu}=-T^{\nu}_{\mu;\nu}
\end{equation}
i.e.,
\begin{equation} 
\Theta^{0}_{\mu,0}+\Theta^{j}_{\mu,j}+\Gamma^{j}_{j0}\Theta_{\mu}^{0}-
\Gamma^{l}_{0\mu}\Theta_{l}^{0}-\Gamma^{0}_{l\mu}\Theta_{0}^{l}+
\Gamma^{j}_{jl}\Theta_{\mu}^{l}-\Gamma^{l}_{j\mu}\Theta_{l}^{j}=-T^{\nu}_{\mu;\nu}
\end{equation}
We obtain for $\mu=i$
\begin{equation} 
[\Theta^{0}_{i,0}+\Gamma^{j}_{j0}\Theta_{i}^{0}-
\Gamma^{l}_{0i}\Theta_{l}^{0}-\Gamma^{0}_{li}\Theta_{0}^{l}]+[\Theta^{j}_{i,j}+
\Gamma^{j}_{jl}\Theta_{i}^{l}-\Gamma^{l}_{ji}\Theta_{l}^{j}]=-T^{\nu}_{i;\nu}
\end{equation}
and for $\mu=0$
\begin{equation} 
[\Theta^{0}_{0,0}+\Gamma^{j}_{j0}\Theta^{0}_{0}]+
[\Theta^{j}_{0,j}+\Gamma^{j}_{jl}\Theta^{l}_{0}]+
[-\Gamma^{l}_{j0}\Theta^{j}_{l}]=-T^{\nu}_{0;\nu}
\end{equation}

(5.2.3) is the system of dynamical equations for the momentum compenson $\Theta^{0}_{i}$, and (5.2.4) is a dynamical equation for the energy compenson $\Theta^{0}_{0}$.

\subsection{Dynamical equations for the stress compenson}

To establish dynamical equations for the stress compenson, we exploit the space metric $(-g_{ij})$ and the related Christoffel symbols
\begin{equation} 
\Gamma^{(-)}{}^{l}_{ij}=\Gamma^{l}_{ij}
\end{equation}
In the capacity of the equations, we introduce wave equations:
\begin{equation} 
\ddot{\Theta}^{j}_{i}+g^{mn}\Theta^{j}_{i;m;n}=0
\end{equation}
where $\dot{}=\partial/\partial t$.

(5.3.2) is the system of dynamical equations for the stress compenson $\Theta^{j}_{i}$.

\subsection{Invariance properties and reduced compensons}

(5.2.1) is invariant under the change
\begin{equation} 
\Theta_{\mu}^{\nu}\rightarrow\Theta_{\mu}^{\nu}+c_{\Lambda}g^{\nu}_{\mu}\,,\quad
 c_{\Lambda}=\mathrm{const}
\end{equation}
so we put
\begin{equation} 
\Theta_{\mu}^{\nu}=c_{\Lambda}g^{\nu}_{\mu}+\tilde{\Theta}_{\mu}^{\nu}
\end{equation}

Again, (5.3.2) is invariant under the change
\begin{equation} 
\Theta_{i}^{j}\rightarrow\Theta_{i}^{j}+(c_{0}-c_{1}t)g^{j}_{i}\,,\quad
 c_{0},c_{1}=\mathrm{const}
\end{equation}
and we put
\begin{equation} 
\Theta_{i}^{j}=(c_{0}-c_{1}t)g^{j}_{i}+\bar{\Theta}_{i}^{j}
\end{equation}
Thus,
\begin{equation} 
\Theta^{\nu}_{\mu}=c_{\Lambda}g_{\mu}^{\nu}+\delta^{m}_{\mu}\delta^{\nu}_{n}
(c_{0}-c_{1}t)g^{n}_{m}+\bar{\Theta}^{\nu}_{\mu}
\end{equation}
where $\bar{\Theta}^{\nu}_{\mu}$ is the reduced compenson.

Now equations (5.3.2), (5.2.3), (5.2.4) reduce to
\begin{equation} 
\ddot{\bar{\Theta}}^{j}_{i}+g^{mn}\bar{\Theta}^{j}_{i;m;n}=0
\end{equation}
\begin{equation} 
[\dot{\bar{\Theta}}^{0}_{i}+\Gamma^{j}_{j0}\bar{\Theta}_{i}^{0}-
\Gamma^{l}_{0i}\bar{\Theta}_{l}^{0}-\Gamma^{0}_{li}\bar{\Theta}_{0}^{l}]+[\bar{\Theta}^{j}_{i,j}+
\Gamma^{j}_{jl}\bar{\Theta}_{i}^{l}-\Gamma^{l}_{ji}\bar{\Theta}_{l}^{j}]=-T^{\nu}_{i;\nu}
\end{equation}
\begin{equation} 
[\dot{\bar{\Theta}}^{0}_{0,0}+\Gamma^{j}_{j0}\bar{\Theta}^{0}_{0}]+
[\bar{\Theta}^{j}_{0,j}+\Gamma^{j}_{jl}\bar{\Theta}^{l}_{0}]+
[-\Gamma^{j}_{j0}(c_{0}-c_{1}t)-\Gamma^{l}_{j0}\bar{\Theta}^{j}_{l}]=-T^{\nu}_{0;\nu}
\end{equation}
respectively.

The equations for the reduced stress, momentum and energy compensons are (5.4.6), (5.4.7) and (5.4.8), respectively.

Initial data are
\begin{equation} 
\bar{\Theta}^{\nu}_{\mu}(x^{l},0),\quad \dot{\bar{\Theta}}^{j}_{i}(x^{l},0)
\end{equation}

\subsection{The extended Einstein equations}

The extended Einstein equations take the form
\begin{equation} 
G_{i}^{j}-(\Lambda+8\pi\varkappa c_{\Lambda})g_{i}^{j}=8\pi\varkappa[T_{i}^{j}+
(c_{0}-c_{1}t)g_{i}^{j}+\bar{\Theta}_{i}^{j}]
\end{equation}
\begin{equation} 
G_{\mu}^{0}-(\Lambda+8\pi\varkappa c_{\Lambda})g_{\mu}^{0}=8\pi\varkappa[T_{\mu}^{0}+
\bar{\Theta}_{\mu}^{0}]
\end{equation}

\subsection{Cosmological and pressure compensons.\\ Dark energy and dark matter}

In the extended Einstein equations (5.5.1), (5.5.2),
\begin{equation} 
\Theta_{\mathrm{cosmological}}{}_{\,\mu}^{\,\nu}:=8\pi\varkappa c_{\Lambda}g_{\mu}^{\nu}
\end{equation}
is the cosmological compenson and
\begin{equation} 
\Theta_{\mathrm{pressure}}{}_{\,i}^{\,j}:=(c_{0}-c_{1}t)g_{i}^{j}
\end{equation}
is the pressure compenson.

We might put $\Lambda=0$ in (1.2.1) and then put
\begin{equation} 
8\pi\varkappa(\varrho_{\mathrm{vacuum}}+c_{\Lambda})=\Lambda
\end{equation}
where $\varrho_{\mathrm{vacuum}}$ is the vacuum energy density. Thus, $c_{\Lambda}$ gives rise to dark energy.

Next,
\begin{equation} 
\Theta_{\mathrm{pressure}}{}_{\,i}^{\,j}+\bar{\Theta}_{i}^{j}\quad \mathrm{and}\quad
\bar{\Theta}_{\mu}^{0}
\end{equation}
may be interpreted as the energy-momentum tensor of dark matter.

\subsection{Dynamics structure}

In compensational gravity, it is assumed that the energy-momentum tensor of matter, $T_{\mu}^{\nu}$, is governed by matter dynamics equations. Thus, the dynamical structure of the theory is the following: (5.5.1) and (5.4.6) are 6+6 equations for 6+6 quantities $g_{ij}$ and $\bar{\Theta}_{i}^{j}$; (5.4.7) are 3 equations for 3 quantities $\bar{\Theta}_{i}^{0}$; and (5.4.8) is an equation for $\bar{\Theta}_{0}^{0}$; (5.5.2) at $t=0$ are constraints on initial conditions.

However, $\bar{\Theta}_{\mu}^{0}$ are determined by (5.5.2) for $t>0$, so that, in fact, there is no need in equations (5.4.7), (5.4.8).

\section{Statics}

\subsection{Metric, the Christoffel symbols, tensor components,\\ and the extended Einstein equations}

In statics, metric is of the form
\begin{equation} 
\mathrm{d}s^{2}=g_{00}\mathrm{d}t^{2}+g_{ij}\mathrm{d}x^{i}\mathrm{d}x^{j}\,,\quad
g=g(x^{l})
\end{equation}
The Christoffel symbols meet the conditions [22]
\begin{equation} 
\Gamma^{0}_{00}=0,\quad\Gamma^{0}_{ij}=0,\quad\Gamma^{i}_{0j}=0
\end{equation}
There are identities
\begin{equation} 
A^{i}_{0}=0,\;A^{0}_{i}=0\quad A=G,T,\Theta
\end{equation}
The extended Einstein equations are
\begin{equation} 
G^{0}_{0}-\Lambda=8\pi\varkappa(T^{0}_{0}+\Theta^{0}_{0})
\end{equation}
\begin{equation} 
G^{j}_{i}-\Lambda g^{j}_{i}=8\pi\varkappa(T^{j}_{i}+\Theta^{j}_{i})
\end{equation}
The identity (2.4.1) reduces to
\begin{equation} 
(G-\Lambda g)^{j}_{i;j}\equiv 0
\end{equation}

\subsection{Equations for the compenson}

Equation (5.2.1)
\begin{equation} 
\Theta_{\mu}^{\nu}{}_{;\nu}=-T_{\mu}^{\nu}{}_{;\nu}
\end{equation}
reduces to 3 equations:
\begin{equation} 
\Theta_{i,j}^{j}+\Gamma^{\nu}_{j\nu}\Theta_{i}^{j}-\Gamma^{l}_{ij}\Theta_{l}^{j}-
\Gamma^{0}_{i0}\Theta_{0}^{0}=-T^{\nu}_{i;\nu}\,,\quad i=1,2,3
\end{equation}

In consequence of (6.2.2), (6.1.6), in the system of equations (6.1.4), (6.1.5) there are only 4 independent equations. 3 more equations for metric should be introduced. 

\subsection{Invariance properties and reduced compensons}

Equations (6.2.2) are invariant under the change
\begin{equation} 
\Theta_{\mu}^{\nu}\rightarrow \Theta_{\mu}^{\nu}+c_{\Lambda} g_{\mu}^{\nu}
\end{equation}
so we put
\begin{equation} 
\Theta_{\mu}^{\nu}=c_{\Lambda} g_{\mu}^{\nu}+\bar{\Theta}_{\mu}^{\nu}
\end{equation}
and obtain from (6.2.2)
\begin{equation} 
\bar{\Theta}_{i,j}^{j}+\Gamma^{\nu}_{j\nu}\bar{\Theta}_{i}^{j}-
\Gamma^{l}_{ij}\bar{\Theta}_{l}^{j}-
\Gamma^{0}_{i0}\bar{\Theta}_{0}^{0}=-T^{\nu}_{i;\nu}\,,\quad i=1,2,3
\end{equation}
The extended Einstein equations (6.1.4), (6.1.5) take the form
\begin{equation} 
G^{0}_{0}-(\Lambda+8\pi\varkappa c_{\Lambda})=8\pi\varkappa(T^{0}_{0}+
\bar{\Theta}^{0}_{0})
\end{equation}
\begin{equation} 
G^{j}_{i}-(\Lambda+8\pi\varkappa c_{\Lambda}) g^{j}_{i}=8\pi\varkappa(T^{j}_{i}+\bar{\Theta}^{j}_{i})
\end{equation}

 A static state depends on the prehistory, i.e., a dynamical process resulting in the static state. There are 3 equations (6.3.3) for 7 quantities $\bar{\Theta}^{0}_{0},\,\bar{\Theta}^{j}_{i}$. 4 of those components and components of the metric (6.1.1) may be prescribed in some coordinate system. (This is the approach pointed out by Kretschmann [26].)

\section{Purely dark objects}

\subsection{Definition}

A purely dark object is defined by the following conditions:
\begin{equation} 
T_{\mu}^{\nu}=0
\end{equation}
\begin{equation} 
\Theta_{\mu}^{\nu}\neq 0
\end{equation}

The characteristic features of a purely dark object are these: The object is transparent for light and material bodies and, with the exception of compensation, may be well apart from material objects.

\subsection{Do purely dark objects exist?}

If the only predestination of the compenson consists in compensating for the insufficiency of the energy-momentum tensor in the extended Einstein equation, then purely dark objects should not exist.

\section{The cycling universe}

\subsection{Setting of a problem}

We set up the problem of a cycling closed universe by introducing the radius of the universe, $R(t)$ [27]. Cosmic space for the closed universe is a three-sphere, $S^{3}$, with the space volume
\begin{equation} 
V(t)=\int\limits_{S^{3}}|-g|\mathrm{d}^{3}x\,\quad g=\mathrm{det}(g_{ij})
\end{equation}
Put
\begin{equation} 
R:=(V/2\pi^{2})^{1/3}\,,\quad R=R(t)
\end{equation}
and represent the metric (5.1.1) in the form
\begin{equation} 
\mathrm{d}s^{2}=\mathrm{d}t^{2}-R^{2}(t)h_{ij}\mathrm{d}x^{i}\mathrm{d}x^{j}
\end{equation}

Now the problem is reduced to finding the character of the time dependence of $R$. To obtain cyclicity, it is necessary to avoid two singularities: $R=0$ and $R\rightarrow\infty$.

In this Section we carry out a qualitative analysis, and in the next Section we treat the compensational FLRW universe.

\subsection{Deflation-inflation}

The singularity $R=0$ is avoided due to the deflation-inflation process [28], in which $R$ passes through a minimum:
\begin{equation} 
R_{\mathrm{min}}=R(t_{\mathrm{min}})>0,\quad \dot{R}(t_{\mathrm{min}})=0,\quad
\ddot{R}(t_{\mathrm{min}})>0
\end{equation}

\subsection{Expansion-contraction}

The singularity $R\rightarrow\infty$ is avoided due to the pressure compenson
\begin{equation} 
\Theta_{\mathrm{pressure}}{}_{\,i}^{\,j}=(c_{0}-c_{1}t)g_{i}^{j}
\end{equation}
in (5.5.1). With $c_{1}>0$, $R$ passes through a maximum:
\begin{equation} 
R_{\mathrm{max}}=R(t_{\mathrm{max}})<\infty,\quad \dot{R}(t_{\mathrm{max}})=0,\quad
\ddot{R}(t_{\mathrm{max}})<0
\end{equation}

\subsection{Cycle}

Introduce
\begin{equation} 
t_{\mathrm{min}\,n}<t_{\mathrm{max}\,n}<t_{\mathrm{min}\,n+1}\,,\quad
 n=0,\mp 1,\mp 2,\cdots
\end{equation}
Put
\begin{equation} 
\mathrm{for}\;t_{\mathrm{min}\,n}<t<t_{\mathrm{max}\,n}\quad
\Theta_{\mathrm{pressure}}{}_{\,1}^{\,1}=c_{0}-c_{1}(t-t_{\mathrm{min}\,n})
\end{equation}
and
\begin{equation} 
\mathrm{for}\;t_{\mathrm{max}\,n}<t<t_{\mathrm{min}\,n+1}\quad
\Theta_{\mathrm{pressure}}{}_{\,1}^{\,1}=c_{0}-c_{1}(t_{\mathrm{max}\,n}
-t_{\mathrm{min}\,n})+c_{1}(t-t_{\mathrm{max}\,n})
\end{equation}
Thus,
\begin{equation} 
\Theta_{\mathrm{pressure}}{}_{\,1}^{\,1}(t_{\mathrm{min}\,n}+0)=c_{0}
\end{equation}
\begin{equation} 
\Theta_{\mathrm{pressure}}{}_{\,1}^{\,1}(t_{\mathrm{max}\,n})=
c_{0}-c_{1}(t_{\mathrm{max}\,n}
-t_{\mathrm{min}\,n})
\end{equation}
\begin{equation} 
\Theta_{\mathrm{pressure}}{}_{\,1}^{\,1}(t_{\mathrm{min}\,n+1}-0)=
c_{0}-c_{1}[(t_{\mathrm{max}\,n}
-t_{\mathrm{min}\,n})-(t_{\mathrm{min}\,n+1}-t_{\mathrm{max}\,n})]
\end{equation}
\begin{equation} 
\Theta_{\mathrm{pressure}}{}_{\,1}^{\,1}(t_{\mathrm{min}\,n}-0)=
c_{0}-c_{1}[(t_{\mathrm{max}\,n-1}
-t_{\mathrm{min}\,n-1})-(t_{\mathrm{min}\,n}-t_{\mathrm{max}\,n-1})]
\end{equation}
Generally,
\begin{equation} 
\Theta_{\mathrm{pressure}}{}_{\,1}^{\,1}(t_{\mathrm{min}\,n}+0)\neq
\Theta_{\mathrm{pressure}}{}_{\,1}^{\,1}(t_{\mathrm{min}\,n}-0)
\end{equation}

A cycle corresponds to times
\begin{equation} 
t_{\mathrm{min}\,n}<t<t_{\mathrm{min}\,n+1}\,,\quad n=\cdots-2,-1,0,1,2,\cdots
\end{equation}

\subsection{On the aging problem}

For the cycling, or oscillating model, there exists the aging problem [11]:``In each cycle the ratio of photons to nuclear particles \ldots is slightly increased by a kind of friction \ldots so it is hard to see how the universe could have previously experienced an infinite number of cycles.''

To circumvent the problem, we assume that, after the end of each deflation-inflation phase, the state of the universe is, on the average, the same.

\section{The cycling compensational FLRW universe}

\subsection{Isotropy}

In the case of the closed isotropic universe, metric is of the form [29]
\begin{equation} 
\mathrm{d}s^{2}=\mathrm{d}t^{2}-R^{2}(t)\left[\frac{\mathrm{d}r^{2}}{1-r^{2}}+
r^{2}(\mathrm{d}\theta^{2}+\sin^{2}\theta\, \mathrm{d}\phi^{2})\right]
\end{equation}
and the Christoffel symbols are
\begin{eqnarray}
\Gamma_{11}^{0}&=&R\dot{R}/(1-r^{2})\qquad\Gamma_{22}^{0}=r^{2}R\dot{R}\qquad
\Gamma_{33}^{0}=\Gamma_{22}^{0}\sin^{2}\theta
\nonumber\\
\Gamma_{01}^{1}&=&\dot{R}/R\qquad\Gamma_{11}^{1}=r/(1-r^{2})
\qquad\Gamma_{22}^{1}=-r(1-r^{2})\qquad
\Gamma_{33}^{1}=\Gamma_{22}^{1}\sin^{2}\theta\nonumber\\
\Gamma_{02}^{2}&=&\Gamma_{03}^{3}=\dot{R}/R\qquad\Gamma_{12}^{2}=\Gamma_{13}^{3}=1/r
\nonumber\\
\Gamma_{33}^{2}&=&-\sin\theta\cos\theta\qquad\Gamma_{23}^{3}=\cot\theta\qquad
\end{eqnarray}
Nonzero tensor components are
\begin{equation} 
A_{0}^{0},\;A_{1}^{1}=A_{2}^{2}=A_{3}^{3},\quad A=G,T,\Theta,\quad A=A(t)
\end{equation}

\subsection{Compenson}

From
\begin{equation} 
\bar{\Theta}^{j}_{i;m}=\bar{\Theta}^{j}_{i,m}+\Gamma^{j}_{ml}\bar{\Theta}_{i}^{l}-
\Gamma^{l}_{mi}\bar{\Theta}_{l}^{j}
\end{equation}
follows
\begin{equation} 
\bar{\Theta}^{1}_{1;m}=0
\end{equation}
Thus,
\begin{equation} 
\ddot{\bar{\Theta}}^{1}_{1}=0
\end{equation}
and, in view of (5.4.4),
\begin{equation} 
\bar{\Theta}^{1}_{1}=0
\end{equation}

Next, (5.4.8) reduces to
\begin{equation} 
\dot{\bar{\Theta}}^{0}_{0}+3\frac{\dot{R}}{R}\bar{\Theta}^{0}_{0}-
3\frac{\dot{R}}{R}(c_{0}-c_{1}t)=-[\dot{T}^{0}_{0}+3\frac{\dot{R}}{R}T^{0}_{0}-
3\frac{\dot{R}}{R}T^{1}_{1}]
\end{equation}
or
\begin{equation} 
\frac{\mathrm{d}}{\mathrm{d}t}[(\bar{\Theta}+T^{0}_{0})R^{3}]-
\frac{\mathrm{d}R^{3}}{\mathrm{d}t}[T^{1}_{1}+(c_{0}-c_{1}t)]
\end{equation}
hence
\begin{equation} 
\bar{\Theta}^{0}_{0}=-T^{0}_{0}+\frac{c_{2}}{R^{3}}+\frac{1}{R^{3}}
\int\limits_{0}^{t}\frac{\mathrm{d}R^{3}}{\mathrm{d}t}T^{1}_{1}\mathrm{d}t+
c_{0}-c_{1}t+c_{1}\frac{1}{R^{3}}\int\limits_{0}^{t}R^{3}\mathrm{d}t
\end{equation}

\subsection{The extended Einstein equations}

Equations (5.5.1), (5.5.2) reduce to
\begin{equation} 
2\frac{\ddot{R}}{R}+\frac{\dot{R}^{2}}{R}+\frac{1}{R^{2}}-(\Lambda+
8\pi\varkappa c_{\Lambda})=8\pi\varkappa[T^{1}_{1}+(c_{0}-c_{1}t)]
\end{equation}
\begin{equation} 
3\left(\frac{\dot{R}^{2}}{R^{2}}+\frac{1}{R^{2}}\right)-(\Lambda+
8\pi\varkappa c_{\Lambda})=8\pi\varkappa[T^{0}_{0}+\bar{\Theta}^{0}_{0}]
\end{equation}
From (9.3.2) follows another expression for $\bar{\Theta}^{0}_{0}$:
\begin{equation} 
\bar{\Theta}^{0}_{0}=-T^{0}_{0}+\frac{1}{8\pi\varkappa}
\left[3\left(\frac{\dot{R}^{2}}{R^{2}}+\frac{1}{R^{2}}\right)-
(\Lambda+
8\pi\varkappa c_{\Lambda})\right]
\end{equation}

\subsection{Large radius}

Let the radius $R$ be large so that the terms $1/R^{2}$ and $8\pi\varkappa T^{1}_{1}$ in (9.3.1) are small and (9.3.1) reduces to
\begin{equation} 
2\frac{\ddot{R}}{R}+\frac{\dot{R}^{2}}{R}=a_{0}-a_{1}t
\end{equation}
where
\begin{equation} 
a_{0}=\Lambda+8\pi\varkappa(c_{\Lambda}+c_{0})\,,\quad a_{1}=8\pi\varkappa c_{1}
\end{equation}

Put
\begin{equation} 
R(t)=R_{1}\mathrm{e}^{f(t)}
\end{equation}
Then
\begin{equation} 
\dot{R}=\dot{f}R,\quad \ddot{R}=(\ddot{f}+\dot{f}^{2})R
\end{equation}
and the equation for $f$ is
\begin{equation} 
\ddot{f}+\frac{3}{2}\dot{f}^{2}=\frac{a_{0}}{2}-\frac{a_{1}}{2}t
\end{equation}
Introduce a dimensionless time $\tau$:
\begin{equation} 
t=\beta\tau,\quad\beta=\left(\frac{2}{a_{1}}\right)^{1/3}
\end{equation}
Then (9.4.5) takes the form
\begin{equation} 
[f''+(f')^{2}]+\frac{1}{2}(f')^{2}=\tau_{0}-\tau
\end{equation}
where $'=\mathrm{d/d} \tau\mathrm{\mathrm{}}$,
\begin{equation} 
\tau_{0}=\frac{\beta^{2}a_{0}}{2}=\frac{a_{0}}{2}\left(\frac{2}{a_{1}}\right)^{2/3}
\end{equation}
From (9.4.4) follows
\begin{equation} 
R'=f'R,\quad R''=[f''+(f')^{2}]R
\end{equation}

We assume that
\begin{equation} 
\tau_{0}\gg 1
\end{equation}
Now,
\begin{equation} 
f''<0\quad \mathrm{for}\;\tau>\tau_{0}
\end{equation}
\begin{equation} 
f''+(f')^{2}=0,\;\;R''=0\quad \mathrm{at}\;\tau=\tau^{(2)}<\tau_{0}
\end{equation}
\begin{equation} 
f'=0,\;\;R'=0\quad \mathrm{at}\; \tau=\tau^{(1)}>\tau_{0}
\end{equation}
\begin{equation} 
f''<0,\;\;R''<0\quad \mathrm{at}\; \tau=\tau^{(1)}
\end{equation}
and
\begin{equation} 
1\ll \tau^{(2)}<\tau_{0}<\tau^{(1)}
\end{equation}

Next, put
\begin{equation} 
f'=y,\quad x=\tau-\tau_{0}\,,\quad '=\mathrm{d/d}\tau=\mathrm{d/d}x
\end{equation}
then
\begin{equation} 
y'+\frac{3}{2}y^{2}+x=0
\end{equation}
and
\begin{equation} 
y=0 \quad \mathrm{at}\; x^{1}=\tau^{1}-\tau_{0}>0
\end{equation}

An exact solution to equation (9.4.17) is expressed in terms of special functions, but it is rather complicated.

Consider a neighborhood of the point $x=0$:
\begin{equation} 
y=\alpha_{0}+\alpha_{1}x+\alpha_{2}x^{2}+\cdots,\quad\alpha_{0}>0
\end{equation}
We find
\begin{equation} 
y'+y^{2}=-\frac{1}{2}\alpha_{0}^{2}+\left(\frac{3}{2}\alpha_{0}^{3}-1\right)x+\cdots
\end{equation}
Thus,
\begin{equation} 
\mathrm{at}\; x=0\quad y'+y^{2}<0,\;\;R''<0
\end{equation}
which corresponds to (9.4.1):
\begin{equation} 
\mathrm{for}\;a_{0}-a_{1}t=0\;\mathrm{and}\;\dot{R}\neq 0,\quad\ddot{R}<0
\end{equation}

\subsection{Cycles}

All cycles are the same. Put
\begin{equation} 
t_{\mathrm{min}\,0}=0
\end{equation}
then
\begin{equation} 
\mathrm{for}\;t_{\mathrm{min}\,0}<t<t_{\mathrm{max}\,0}\quad
\Theta_{\mathrm{pressure}}{}_{\,1}^{\,1}=c_{0}-c_{1}t
\end{equation}
\begin{equation} 
\mathrm{for}\;t_{\mathrm{max}\,-1}<t<t_{\mathrm{min}\,0}\quad
\Theta_{\mathrm{pressure}}{}_{\,1}^{\,1}=c_{0}-c_{1}(-t)
\end{equation}
Now,
\begin{equation} 
t_{\mathrm{max}\,n}-t_{\mathrm{min}\,n}=t_{\mathrm{min}\,n+1}-t_{\mathrm{max}\,n}
\quad\mathrm{is\;independent\;of}\;n
\end{equation}
\begin{equation} 
\Theta_{\mathrm{pressure}}{}_{\,1}^{\,1}(t_{\mathrm{min}\,n}+0)=
\Theta_{\mathrm{pressure}}{}_{\,1}^{\,1}(t_{\mathrm{min}\,n}-0)=
\Theta_{\mathrm{pressure}}{}_{\,1}^{\,1}(t_{\mathrm{min}\,n})=c_{0}
\end{equation}

A cycle is
\begin{equation} 
t_{\mathrm{min}\,n}<t<t_{\mathrm{min}\,n+1}
\end{equation}
and its period
\begin{equation} 
t_{\mathrm{min}\,n+1}-t_{\mathrm{min}\,n}=2(t_{\mathrm{max}\,n}-t_{\mathrm{min}\,n})
\end{equation}
does not depend on $n$.

\section{Cycle phases}

Cycle phases are these:

$R_{\mathrm{min}\,n}$, inflation (scalar field)

expansion (ordinary matter)

exponential expansion ($\Lambda$)

expansion, $R_{\mathrm{max}\,n}$, contraction (pressure compenson)

exponential contraction ($\Lambda$)

contraction (ordinary matter)

deflation (scalar field), $R_{\mathrm{min}\,n+1}$

\noindent In parentheses, the governing factor is indicated.
\newpage
\section*{Acknowledgments}

I would like to thank Alex A. Lisyansky for support and Stefan V.
Mashkevich for helpful discussions.

\end{document}